\documentclass[a4paper,fleqn,usenatbib]{mnras}

\usepackage{newtxtext,newtxmath}

\usepackage[T1]{fontenc}
\usepackage{ae,aecompl}

\DeclareRobustCommand{\VAN}[3]{#2}
\let\VANthebibliography\thebibliography
\def\thebibliography{\DeclareRobustCommand{\VAN}[3]{##3}\VANthebibliography}

\usepackage{graphicx}	
\usepackage{amsmath}	

\usepackage{grffile}
\usepackage{braket}

\usepackage{bm} 
\usepackage{ulem}

\title[Wide-angle effects]{Wide-angle effects on galaxy ellipticity correlations}

\author[M. Shiraishi et al.]{Maresuke Shiraishi,$^{1}$\thanks{E-mail: shiraishi-m@t.kagawa-nct.ac.jp}
Atsushi Taruya,$^{2,4}$
Teppei Okumura,$^{3,4}$
Kazuyuki Akitsu$^{4}$
\\
$^1$Department of General Education, National Institute of Technology, Kagawa College, 355 Chokushi-cho, Takamatsu, Kagawa 761-8058, Japan \\
$^2$Center for Gravitational Physics, Yukawa Institute for Theoretical Physics, Kyoto University, Kyoto 606-8502, Japan \\
$^3$Academia Sinica Institute of Astronomy and Astrophysics (ASIAA), No. 1, Section 4, Roosevelt Road, Taipei 10617, Taiwan \\
$^4$Kavli Institute for the Physics and Mathematics of the Universe (WPI), UTIAS, The University of Tokyo, Kashiwa, Chiba 277-8583, Japan 
}

\date{Accepted XXX. Received YYY; in original form ZZZ}

\pubyear{2020}

\begin{document}
\label{firstpage}
\pagerange{\pageref{firstpage}--\pageref{lastpage}}
\maketitle

\begin{abstract}

  We show an efficient way to compute wide-angle or all-sky statistics of galaxy intrinsic alignment in three-dimensional configuration space. For this purpose, we expand the two-point correlation function using a newly introduced spin-dependent tripolar spherical harmonic basis. Therefore, the angular dependences on the two line-of-sight (LOS) directions pointing to each pair of objects, which are degenerate with each other in the conventional analysis under the small-angle or plane-parallel (PP) approximation, are unambiguously decomposed. By means of this, we, for the first time, compute the wide-angle auto and cross correlations between intrinsic ellipticities, number densities and velocities of galaxies, and compare them with the PP-limit results. For the ellipticity-ellipticity and density-ellipticity correlations, we find more than $10\%$ deviation from the PP-limit results if the opening angle between two LOS directions exceeds $30^\circ - 50^\circ$. It is also shown that even if the PP-limit result is strictly zero, the non-vanishing correlation is obtained over the various scales, arising purely from the curved-sky effects. Our results indicate the importance of the data analysis not relying on the PP approximation in order to determine the cosmological parameters more precisely and/or find new physics via ongoing and forthcoming wide-angle galaxy surveys.
\end{abstract}

\begin{keywords}
gravitational lensing: weak -- cosmology: observations -- cosmology: theory -- dark energy -- dark matter -- large-scale structure of Universe.
\end{keywords}


\section{Introduction}

The large-scale structure of the Universe traced by the spatial distribution of galaxies provides a wealth of cosmological information. In particular, its statistical properties, quantified by the two-point correlation function of number densities and peculiar velocities of galaxies, have been playing a major role to test and constrain the cosmology. Besides, the information on individual shapes and orientations of galaxy images, previously treated as a contaminant in weak lensing data analyses \citep{Heavens:2000ad,Croft:2000gz,Crittenden:2000au,Hirata:2004gc,Mandelbaum:2006,Hirata:2007,Okumura:2009,Okumura:2009a}, has recently attracted much attention, and is considered as a beneficial cosmological probe. There are numerous works developed on the formalism and theoretical predictions of the galaxy intrinsic alignment (IA) statistics not only in the projected sky defined on two-dimensional celestial sphere \citep{Catelan:2000vm,Heavens:2000ad,Hirata:2004gc,Schmidt:2015xka,Kogai:2018nse,Biagetti:2020lpx,Kogai:2020vzz,Vlah:2020ovg}, but also in the three-dimensional configuration or Fourier space \citep{Crittenden:2000au,Schmidt:2012ne,Okumura:2019ozd,Vlah:2019byq,Okumura:2019ned,Okumura:2020a,Akitsu:2020fpg}. Moreover, several studies have forecasted the constraints on key cosmological parameters such as the dark energy equation-of-state parameters and the parameters related to the early Universe, and found a substantial improvement on their precision and/or detectability when combining the IA statistics \citep{Chisari:2013dda,Schmidt:2015xka,Chisari:2016xki,Kogai:2018nse,Taruya:2020tdi,Akitsu:2020jvx}.

Motivated by these, this Letter examines an uninvestigated and important issue on the three-dimensional two-point correlation of IA for a widely separated pair, namely the wide-angle effects. The correlation function for a pair of target fields at positions ${\bf x}_1$ and ${\bf x}_2$ is generally characterized by the three vectors: two line-of-sight (LOS) directions $\hat{x}_{1}$ and $\hat{x}_2$ and the separation vector ${\bf x}_{12} \equiv {\bf x}_1 - {\bf x}_2$. On the other hand, the so-called plane-parallel (PP) approximation, which sets $\hat{x}_1 = \hat{x}_2$, has been frequently imposed in the previous analyses. While this greatly simplifies the computation of correlation function, the accuracy of the approximation is no longer ensured when the opening angle between $\hat{x}_{1}$ and $\hat{x}_2$, dubbed as $\Theta$, becomes large enough. Indeed, in the correlation functions of galaxy number density and velocity fields, more than $10\%$ accuracy loss occurs for $\Theta$ of a few tens of degrees \citep{Szapudi:2004gh,Yoo:2013zga,Castorina:2017inr,Taruya:2019xsf,Castorina:2019hyr}. In anticipation of upcoming wide-angle or all-sky galaxy IA surveys, the development of the comprehensive formalism without the PP approximation is thus timely and crucial.

To cope with this, we perform the decomposition using the tripolar spherical harmonic (TripoSH) basis, i.e., the irreducible tensor products of the spherical harmonics with different arguments \citep{Varshalovich:1988ye}, and accordingly resolve tangled angular dependence between $\hat{x}_1$, $\hat{x}_2$ and $\hat{x}_{12}$ in the correlation function. Then, to treat the correlation functions of the spin-2 intrinsic ellipticity field, we implement a new basis defined by adding the spin dependence to the conventional spin-0 TripoSH basis utilized for the correlations of the density and velocity fields \citep{Szapudi:2004gh,Yoo:2013zga,Shiraishi:2020nnw}. Based on this decomposition, the resultant TripoSH coefficients are obtained in a rather simplified and computable form, involving only the one-dimensional integral.

With numerics of the TripoSH coefficients, we compute the ellipticity auto correlation and the density-ellipticity and velocity-ellipticity cross correlations. In comparison with the results obtained under the PP approximation, we find more than $10\%$ deviation for $\Theta \gtrsim 30^\circ - 50^\circ$ in the ellipticity-ellipticity and density-ellipticity correlations. Regarding the velocity-ellipticity correlation, even if the PP-limit result is strictly zero, the non-vanishing correlation is found to be realized over the various scales.

\section{Statistics of intrinsic alignment, number density and peculiar velocity}

\begin{figure*}
  \includegraphics[width = 1.\textwidth]{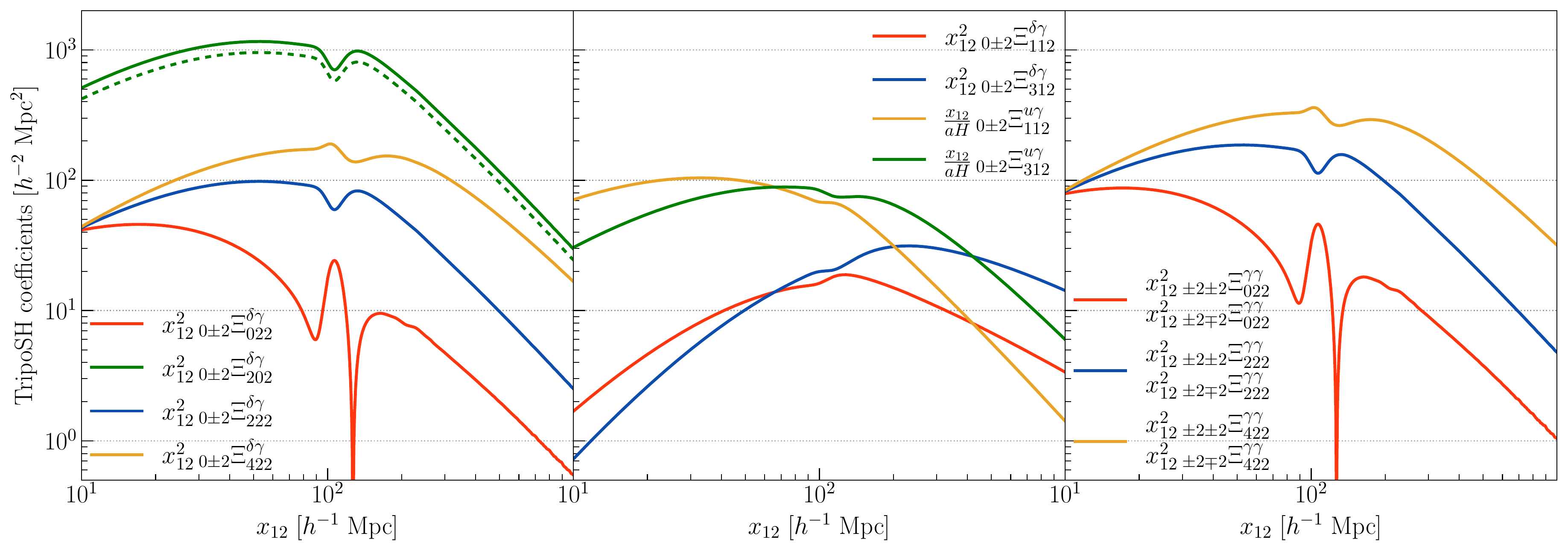}
  \caption{Scale dependence of all non-vanishing TripoSH coefficients for the ellipticity correlation functions, ${}_{0 \pm 2}\Xi_{\ell \ell_1 \ell_2}^{\delta \gamma}$ (left panel for $\ell = \rm even$ and center panel for $\ell = \rm odd$), ${}_{0 \pm 2}\Xi_{\ell \ell_1 \ell_2}^{u \gamma}$ (center panel), ${}_{\pm 2 \pm 2}\Xi_{\ell \ell_1 \ell_2}^{\gamma \gamma}$ and ${}_{\pm 2 \mp 2}\Xi_{\ell \ell_1 \ell_2}^{\gamma \gamma}$ (right panel). Taking their absolute values, the results multiplied by $x_{12}^2$ or $x_{12}/(aH)$ are plotted as function of separation, $x_{12}$. Here, we specifically consider the case with $z_1 = z_2 = 0.3$ and $b_{\rm g} = b_{\rm K} = 1$. For the $\delta \gamma$ correlation, distinct results are obtained in real and redshift space, and their non-vanishing coefficients are depicted as dashed and solid lines, respectively. Note that all the coefficients shown here are invariant under the sign inversion of helicities [see Eq.~\eqref{eq:TripoSH_Pi_delta_u_lam_iso}]}\label{fig:xi_TripoSHcoeff_per_x12}  
\end{figure*}

Since our main interests lie in large-scale galaxy correlations, we are allowed to quantify the intrinsic ellipticities, number densities and LOS peculiar velocities of galaxies on the basis of the linear theory \citep{Hamilton:1997zq,Yoo:2013zga}. According to \citet{Catelan:2000vm,Hirata:2004gc}, the ellipticity field $\gamma_{ij}$, defined by the transverse and traceless projection of the second moment of the surface brightness of galaxies, is linearly connected to the real-space matter fluctuation $\delta_{\rm m}$ through
\begin{align}
  \gamma_{ij}({\bf x}) &= \frac{1}{2} \left[ P_{ik}(\hat{x}) P_{j l}(\hat{x})
+ P_{il}(\hat{x}) P_{jk}(\hat{x})
- P_{ij}(\hat{x}) P_{k l}(\hat{x}) \right] \nonumber \\ 
  &\quad \times  \int \frac{d^3 k}{(2\pi)^3} e^{i {\bf k} \cdot {\bf x}}
  \left(\hat{k}_k \hat{k}_l - \frac{1}{3} \delta_{kl} \right) b_{\rm K} \delta_{\rm m}({\bf k}) , \label{eq:gam_ij}
\end{align}
where $P_{ij}(\hat{x}) \equiv \delta_{ij} - \hat{x}_i \hat{x}_j$ and $\hat{x} \equiv {\bf x} / |{\bf x}|$. Here, the parameter $b_{\rm K}$ and field $\delta_{\rm m}$ implicitly depend on time, redshift or the comoving distance, while it is not clearly stated as an argument for notational convenience. This convention is adapted to all variables henceforth unless the parameter dependence is non-trivial. Because of good compatibility with the later TripoSH decomposition, let us further introduce the spin-2 field as
\begin{align}
{}_{\pm 2}\gamma({\bf x}) \equiv m_{\mp}^i(\hat{x}) m_{\mp}^j(\hat{x}) \gamma_{ij}({\bf x}) ,  \label{eq:gam_lam_def}
\end{align}
where the unit vector $\hat{x}$ and its orthonormal vector ${\bf m}_{\pm}(\hat{x})$ are explicitly given as
\begin{align}
  \hat{x} \equiv \left(
  \begin{matrix}
    \sin\theta \cos\phi \\
    \sin\theta \sin\phi \\
    \cos\theta
  \end{matrix}
  \right)
  , \ \ \
 {\bf m}_{\pm}(\hat{x}) \equiv  \frac{1}{\sqrt{2}}
  \left(
  \begin{matrix}
    \cos\theta \cos\phi \pm i \sin\phi \\
    \cos\theta \sin\phi \mp i \cos\phi \\
    -\sin\theta
  \end{matrix}
  \right) . \label{eq:pol_vec_def}
\end{align}
This is expressed as the linear combination of the conventional $+/\times$ modes as
  \begin{align}
    {}_{\pm 2}\gamma({\bf x}) = \frac{1}{2} \left[ \gamma_+({\bf x}) \pm i \gamma_\times({\bf x}) \right]. \label{eq:gam_lam_vs_gam_+x}
  \end{align}
From Eq.~\eqref{eq:gam_ij}, its linear theory expression is given by
\begin{align}
  {}_{\pm 2}\gamma({\bf x}) = 
  \int \frac{d^3 k}{(2\pi)^3} e^{i {\bf k} \cdot {\bf x}}
  \hat{k}_i \hat{k}_j m^i_{\mp}(\hat{x}) m^j_{\mp}(\hat{x})
  b_{\rm K} \delta_{\rm m}({\bf k}) . 
\end{align}

Likewise, the linear theory expressions for the number density fluctuation $\delta({\bf x}) \equiv n({\bf x}) / \bar{n}(x) - 1$ and the LOS velocity field $u({\bf x}) \equiv {\bf v}({\bf x}) \cdot \hat{x}$ are given by
\begin{align}
  \begin{split}
    \delta({\bf x}) &= \int \frac{d^3 k}{(2\pi)^3} e^{i {\bf k} \cdot {\bf x}}
  \left[ 
b_{\rm g}  - 
i \frac{\alpha}{kx} (\hat{k} \cdot \hat{x}) f  
+ (\hat{k} \cdot \hat{x})^2 f  \right] \delta_{\rm m}({\bf k}) , \\
u({\bf x}) &= \int \frac{d^3 k}{(2\pi)^3} e^{i {\bf k} \cdot {\bf x}} \,
i \frac{aH}{k} (\hat{k} \cdot \hat{x}) f \delta_{\rm m} ({\bf k}) ,
\end{split} \label{eq:delta_u}
\end{align}
where $b_{\rm g}$ is the bias parameter for galaxy number density field, $a$ is the scale factor, $H$ is the Hubble parameter, $f$ is the linear growth rate and $\alpha \equiv d \ln \bar{n}(x) / d \ln x +  2$ is the selection function of the galaxy sample. Here, we consider the density field defined in redshift space. The real-space density field is simply obtained by taking the limit, $f \to 0$. Note that for $u$ or ${}_{\pm 2}\gamma$, there is no distinction between real and redshift space at linear order.

For later convenience, we expand these fields using the spin-weighted spherical harmonics as
\begin{align} 
  {}_{\lambda}X({\bf x}) =
  \int \frac{d^3 k}{(2\pi)^3} e^{i {\bf k} \cdot {\bf x}} 
  \sum_{j\mu}  
    \frac{4\pi c_{j}^{X} (k)}{2j+1} Y_{j \mu}(\hat{k}) {}_{-\lambda} Y_{j \mu}^*(\hat{x})
  \delta_{\rm m}({\bf k})
  , \label{eq:delta_u_lam}
\end{align}
where $X = \{\delta, u, \gamma \}$ and 
\begin{align}
  \begin{split}
    c_\ell^\delta(k) &= \left( b_{\rm g} + \frac{1}{3} f \right) \delta_{\ell,0}
    - i \frac{\alpha}{k x} f \, \delta_{\ell,1}
    + \frac{2}{3} f \, \delta_{\ell, 2} ,\\
    c_\ell^u(k) &= i \frac{aH}{k} f \, \delta_{\ell, 1}, \\
    c_\ell^{\gamma}(k) &= \frac{\sqrt{6}}{3} b_{\rm K} \, \delta_{\ell, 2} .
  \end{split} \label{eq:delta_u_lam_coeff}
\end{align}
The subscript in ${}_{\lambda}X$ represents the spin/helicity dependence of each field; therefore, $\lambda = 0$ for $X = \delta, u$, and $\lambda = \pm 2$ for $X = \gamma$. As for the spin-0 fields, for notational simplicity, let us sometimes omit the subscript $0$ in ${}_{0}X$ as in Eq.~\eqref{eq:delta_u}. To derive Eq.~\eqref{eq:delta_u_lam}, we have utilized the spherical harmonic representation of $\hat{k}$, $\hat{x}$ and ${\bf m}_{\pm}(\hat{x})$ as was derived in \citet{Shiraishi:2010kd}. Note that, for $\lambda = 0$, Eq.~\eqref{eq:delta_u_lam} recovers the Legendre expansion.

In this Letter, the matter fluctuation in real space $\delta_{\rm m}$ is assumed to be statistically homogeneous and isotropic, and hence its power spectrum is defined by
\begin{align}
  \Braket{\delta_{\rm m}({\bf k}_1) \delta_{\rm m}({\bf k}_2)} = (2\pi)^3 \delta^{(3)}({\bf k}_1 + {\bf k}_2) P_{\rm m}(k_1) . \label{eq:Pm}
\end{align}
With Eqs.~\eqref{eq:delta_u_lam} and \eqref{eq:Pm} and the addition theorem of the spherical harmonics, the correlation function $\xi_{\lambda_1 \lambda_2}^{X_1 X_2}({\bf x}_{12}, \hat{x}_1, \hat{x}_2) \equiv \Braket{{}_{\lambda_1} X_1({\bf x}_1) {}_{\lambda_2}X_2({\bf x}_2) }$ is computed as
\begin{align}
\xi_{\lambda_1 \lambda_2}^{X_1 X_2}({\bf x}_{12}, \hat{x}_1, \hat{x}_2)
=  \int \frac{d^3 k}{(2\pi)^3} e^{i {\bf k} \cdot {\bf x}_{12}}
P_{\lambda_1 \lambda_2}^{X_1 X_2}({\bf k}, \hat{x}_1, \hat{x}_2) , \label{eq:xi_delta_u_lam}
\end{align}
where  
\begin{align}
P_{\lambda_1 \lambda_2}^{X_1 X_2}({\bf k}, \hat{x}_1, \hat{x}_2)
&= \sum_{J j_1 j_2} \frac{(4\pi)^2 (-1)^{j_2} h_{J j_1 j_2}^{0~0~0}}{(2j_1 + 1)(2j_2 + 1)}  
 c_{j_1}^{X_1} (k) c_{j_2}^{X_2} (k)   \nonumber \\
  & \quad \times
  \sum_{ \mu \mu_1 \mu_2}  
   \left(
   \begin{matrix}
     J & j_1 & j_2 \\
     \mu & \mu_1 & \mu_2 
   \end{matrix}
   \right)
   Y_{J \mu}^*(\hat{k}) 
   \nonumber \\
  & \quad \times
   {}_{-\lambda_1} Y_{j_1 \mu_1}^*(\hat{x}_1)
   {}_{-\lambda_2} Y_{j_2 \mu_2}^*(\hat{x}_2)
   P_{\rm m}(k). \label{eq:P_delta_u_lam}
\end{align}
Here, the function like the $2\times 3$ matrix represents the Wigner $3j$ symbol, and we define
\begin{align}
h_{l_1 l_2 l_3}^{s_1 s_2 s_3} \equiv \sqrt{\frac{(2 l_1 + 1)(2 l_2 + 1)(2 l_3 + 1)}{ 4 \pi}} \left(\begin{matrix}
  l_1 & l_2 & l_3 \\
  s_1 & s_2 & s_3 
\end{matrix}\right).
\end{align}
Note that $P_{\lambda_1 \lambda_2}^{X_1 X_2}({\bf k}, \hat{x}_1, \hat{x}_2)$ in Eq.~\eqref{eq:xi_delta_u_lam} is not the Fourier counterpart of $\xi_{\lambda_1 \lambda_2}^{X_1 X_2}({\bf x}_{12}, \hat{x}_1, \hat{x}_2)$ because there still remains the position dependence.

For the correspondence of $\xi_{\lambda_1 \lambda_2}^{X_1 X_2}$ to the two-dimensional angular correlation ${}_{\lambda_1 \lambda_2} C_{\ell}^{X_1 X_2}$, see Appendix~\ref{appen:Cl}.

\section{Spin-weighted tripolar spherical harmonic decomposition}

Regarding the spin-0 fields such as $\delta$ and $u$, it has already been shown in the literature \citep{Szalay:1997cc,Szapudi:2004gh,Papai:2008bd,Yoo:2013zga,Taruya:2019xsf,Shiraishi:2020nnw} that the intricate angular dependence in the correlation functions arising from the wide-angle effects can be decomposed by means of the TripoSH basis. In this Letter, extending the usual spin-0 basis \citep{Varshalovich:1988ye,Shiraishi:2020nnw} to the spin-weighted version, we will perform the similar decomposition to the correlation functions including the IA. Here, the spin-weighted TripoSH is defined by
\begin{align}
 {}_{\lambda_1 \lambda_2}{\cal X}_{\ell \ell_1\ell_2}(\hat{x}_{12},\hat{x}_1,\hat{x}_2) 
  &\equiv \{ Y_{\ell}(\hat{x}_{12}) \otimes \{ {}_{\lambda_1} Y_{\ell_1}(\hat{x}_1) \otimes {}_{\lambda_2}Y_{\ell_2}(\hat{x}_2) \}_\ell \}_{00} \nonumber \\
  & = \sum_{m m_1 m_2 } 
 (-1)^{\ell_1 + \ell_2 + \ell} 
  \left( \begin{matrix}
    \ell_1 & \ell_2 & \ell \\
    m_1 & m_2 & m
  \end{matrix}  \right) \nonumber \\ 
  &\quad \times
  Y_{\ell m}(\hat{x}_{12}) {}_{\lambda_1}Y_{\ell_1 m_1}(\hat{x}_1) {}_{\lambda_2}Y_{\ell_2 m_2}(\hat{x}_2) . \label{eq:TripoSH_lam_def}
\end{align}
With this new basis, Eq.~\eqref{eq:xi_delta_u_lam} is expanded in the following form:
\begin{align}
  \xi_{\lambda_1 \lambda_2}^{X_1 X_2}({\bf x}_{12}, \hat{x}_1, \hat{x}_2)
  = \sum_{\ell\ell_1\ell_2} {}_{\lambda_1 \lambda_2}\Xi_{\ell\ell_1\ell_2}^{X_1 X_2}(x_{12}) 
  {}_{\lambda_1 \lambda_2}{\cal X}_{\ell\ell_1\ell_2}(\hat{x}_{12},\hat{x}_1,\hat{x}_2) .
  \label{eq:TripoSH_Xi_lam_def}
\end{align}

In order to derive a simple analytical form of ${}_{\lambda_1 \lambda_2}\Xi_{\ell\ell_1\ell_2}^{X_1 X_2}$ from Eq.~\eqref{eq:xi_delta_u_lam}, we also expand
\begin{align}
  P_{\lambda_1 \lambda_2}^{X_1 X_2}({\bf k}, \hat{x}_1, \hat{x}_2)
  = \sum_{\ell\ell_1\ell_2} {}_{\lambda_1 \lambda_2}\Pi_{\ell\ell_1\ell_2 }^{X_1 X_2}(k)
  {}_{\lambda_1 \lambda_2}{\cal X}_{\ell\ell_1\ell_2}(\hat{k},\hat{x}_1,\hat{x}_2)
  \label{eq:P_lambda1_lambda2_x1_x2}
\end{align}
with the spin-weighted TripoSH ${}_{\lambda_1 \lambda_2}{\cal X}_{\ell\ell_1\ell_2}$ similarly defined in the Fourier space. The coefficients in Eqs.~(\ref{eq:TripoSH_Xi_lam_def}) and (\ref{eq:P_lambda1_lambda2_x1_x2}) are then related through
\begin{align}
  {}_{\lambda_1 \lambda_2}\Xi_{\ell\ell_1\ell_2}^{X_1 X_2}(x_{12}) 
  = i^{\ell} \int_0^\infty \frac{k^2 dk}{2\pi^2}  j_{\ell}(k x_{12})
  {}_{\lambda_1 \lambda_2}\Pi_{\ell \ell_1 \ell_2}^{X_1 X_2}(k) . \label{eq:hankel_delta_u_lam}
\end{align}
In deriving the above, we have expanded $e^{i {\bf k} \cdot {\bf x}_{12}}$ in terms of the spherical harmonics, and performed analytically the $\hat{k}$ integral in Eq.~\eqref{eq:xi_delta_u_lam}. The explicit form of ${}_{\lambda_1 \lambda_2}\Pi_{\ell \ell_1 \ell_2}^{X_1 X_2}$ is obtained via the inverse transformation: 
\begin{align}
  {}_{\lambda_1 \lambda_2}\Pi_{\ell \ell_1 \ell_2}^{X_1 X_2}(k) 
  &=  \int d^2 \hat{k}  \int d^2 \hat{x}_1  \int d^2 \hat{x}_2 
   \nonumber \\ 
   &\quad \times
   P_{\lambda_1 \lambda_2}^{X_1 X_2}({\bf k}, \hat{x}_1, \hat{x}_2)
   {}_{\lambda_1 \lambda_2}{\cal X}_{\ell \ell_1\ell_2}^*(\hat{k},\hat{x}_1,\hat{x}_2). \label{eq:TripoSH_Pi_delta_u_lam_def}
\end{align}
Simplifying the integrals of the (spin-weighted) spherical harmonics and the contractions of the Wigner $3j$ symbols, we finally obtain 
\begin{align}
  {}_{\lambda_1 \lambda_2}\Pi_{\ell \ell_1 \ell_2}^{X_1 X_2}(k) 
  = \frac{(4\pi)^2 (-1)^{\lambda_1 + \lambda_2 + \ell_2} h_{\ell \ell_1 \ell_2}^{0~0~0}}{(2\ell_1 + 1)(2\ell_2 + 1)} 
  c_{\ell_1}^{X_1} (k) c_{\ell_2}^{X_2} (k) P_{\rm m}(k) . 
  \label{eq:TripoSH_Pi_delta_u_lam_iso}
\end{align}
The selection rules in harmonic space, $|\ell_1 - \ell_2| \leq \ell \leq \ell_1 + \ell_2$ and $\ell + \ell_1 + \ell_2 = \rm even$, restrict the number of non-vanishing coefficients. Equation~\eqref{eq:TripoSH_Xi_lam_def} with the analytically expressed coefficients \eqref{eq:hankel_delta_u_lam} and \eqref{eq:TripoSH_Pi_delta_u_lam_iso} is the main result of this Letter.

Figure~\ref{fig:xi_TripoSHcoeff_per_x12} plots the shape of all non-zero TripoSH coefficients, ${}_{0 \pm 2}\Xi_{\ell \ell_1 \ell_2}^{\delta \gamma}$, ${}_{0 \pm 2}\Xi_{\ell \ell_1 \ell_2}^{u \gamma}$, ${}_{\pm 2 \pm 2}\Xi_{\ell \ell_1 \ell_2}^{\gamma \gamma}$ and ${}_{\pm 2 \mp 2}\Xi_{\ell \ell_1 \ell_2}^{\gamma \gamma}$, computed from Eqs.~\eqref{eq:hankel_delta_u_lam} and \eqref{eq:TripoSH_Pi_delta_u_lam_iso}, the results of which are multiplied by $x_{12}^2$ or $x_{12} / (aH)$.%
\footnote{See \citet{Szalay:1997cc,Szapudi:2004gh,Papai:2008bd,Yoo:2013zga,Taruya:2019xsf,Shiraishi:2020nnw,Shiraishi:2020pea} for the analysis of the remaining three coefficients, ${}_{00}\Xi_{\ell\ell_1\ell_2}^{\delta \delta}$, ${}_{00}\Xi_{\ell\ell_1\ell_2}^{\delta u}$ and ${}_{00}\Xi_{\ell\ell_1\ell_2}^{uu}$.}
As expected, the baryon acoustic oscillation bump at $x_{12} \simeq 100 ~ h^{-1} \, \rm Mpc$ is clearly seen in each coefficient. To see the impact of the redshift-space distortions (RSD) on the $\delta\gamma$ correlation, we also plot the real-space signal in dashed line. Since $c_\ell^\delta = b_{\rm g} \delta_{\ell, 0}$ in real space, the only non-vanishing coefficient is ${}_{0 \pm 2}\Xi_{202}^{\delta \gamma}$. The results illustrate how the RSD not only changes the clustering amplitude but also produce anisotropies described by the higher multipoles of the TripoSH coefficients.

Finally, while our formalism is obviously general applicable to the predictions for a largely separated pair, setting the vectors $\hat{x}_1$ and $\hat{x}_2$ to $\hat{x}_1=(0,0,1)=\hat{x}_2$ rigorously reproduces the analytical results known in the PP limit \citep{Okumura:2019ned}.%
\footnote{In the PP limit, the relation between our correlation functions, $\xi_{\lambda_1 \lambda_2}^{X_1 X_2}$, and those computed in \citet{Okumura:2019ned}, denoted by $\xi_{\rm g +}$, $\xi_{v +}$ and $\xi_{\pm}$, is explicitly given as follows:
  $\xi_{\rm g +} =  \xi_{0 +2}^{\delta \gamma} +  \xi{}_{0 -2}^{\delta \gamma}$,
  $\xi_{v +} = \xi_{0 +2}^{u \gamma} + \xi_{0 -2}^{u \gamma}$ and
  $\xi_{\pm} = 2(\xi_{+2 \mp 2}^{\gamma\gamma} + \xi_{-2 \pm 2}^{\gamma\gamma} )$.
}

\section{Wide-angle effects}

\begin{figure}
  \includegraphics[width = \columnwidth]{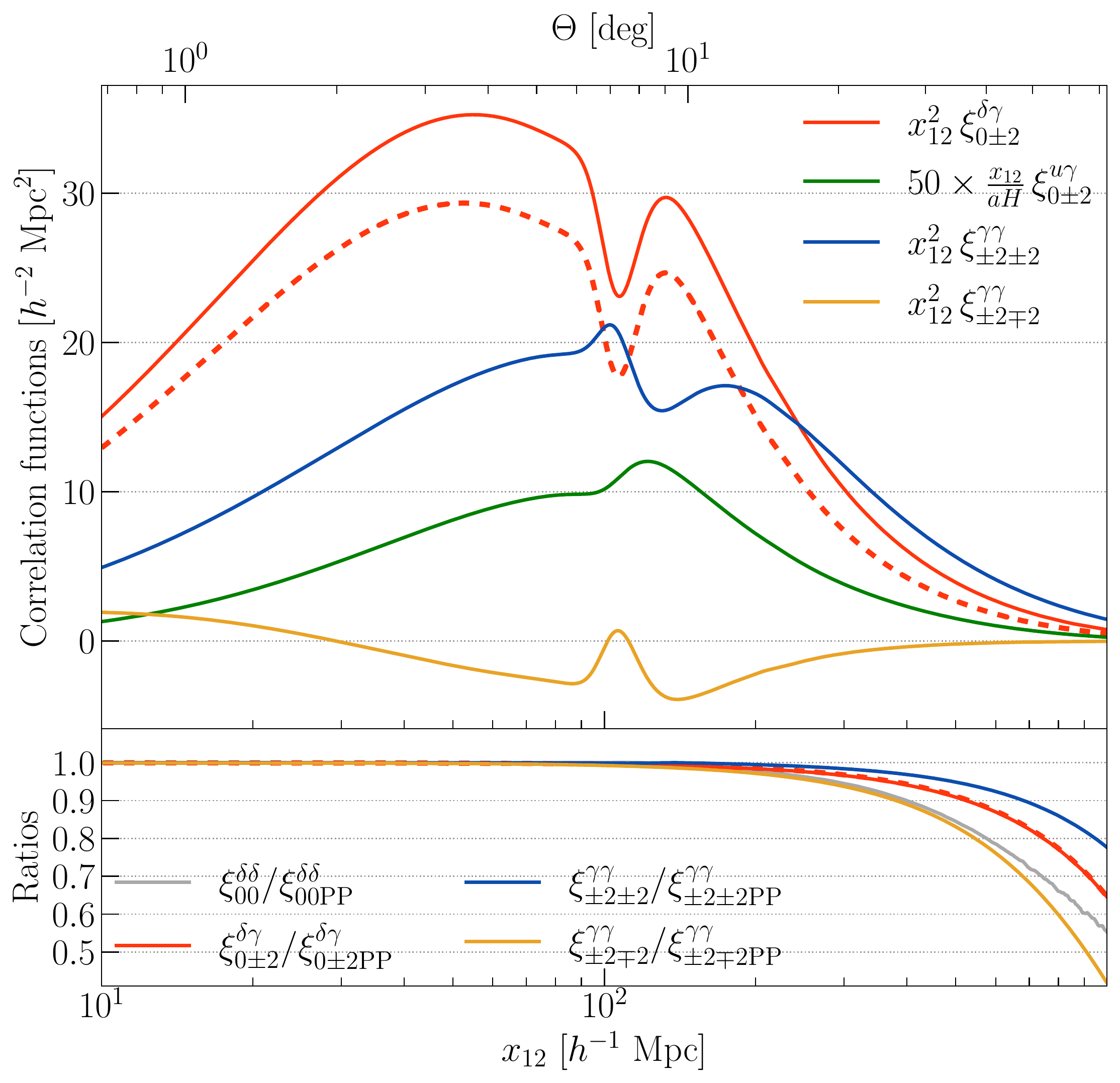}
  \caption{Ellipticity correlation functions, $\xi_{0 \pm 2}^{\delta \gamma}$, $\xi_{0 \pm 2}^{u \gamma}$, $\xi_{\pm 2 \pm 2}^{\gamma \gamma}$ and $\xi_{\pm 2 \mp 2}^{\gamma \gamma}$ (top), and the ratios of these to the corresponding PP-limit results (bottom), plotted against the separation $x_{12}$ or the opening angle $\Theta$ for a specific case with $z_1 = z_2 = 0.3$ and $b_{\rm g} = b_{\rm K} = 1$. Note that the $u \gamma$ correlation in the PP limit strictly vanishes. Hence, in the bottom panel, instead of plotting the $u \gamma$ result, we show the $\delta \delta$ case for a reference. For the $\delta \gamma$ correlation, the results in real and redshift space are shown in dashed and solid lines, respectively.}\label{fig:xi_xiPP_per_x12} 
\end{figure}

Provided the formulas expressed in terms of the spin-weighted TripoSH basis, we are in position to discuss the quantitative impact of the wide-angle effects on the IA statistics. Following \citet{Szapudi:2004gh,Yoo:2013zga}, we consider the coordinate system in which the triangle formed with ${\bf x}_1$, ${\bf x}_2$ and ${\bf x}_{12}$ is confined to the $xy$ plane. Moreover, the separation and direction vectors are set to $\hat{x}_{12} = (1, 0, 0)$, $\hat{x}_1 = (\cos \phi_1 , \sin \phi_1 , 0)$ and $\hat{x}_2 = (\cos \phi_2 , \sin \phi_2 , 0)$ for $\phi_2 \geq \phi_1$. Then the opening angle between $\hat{x}_1$ and $\hat{x}_2$ is defined to be $\Theta \equiv \phi_2 - \phi_1$. The explicit representations of ${}_{\pm 2} \gamma({\bf x}_{1,2})$ and $\gamma_{+ / \times}({\bf x}_{1,2})$ under this coordinate system are obtained by taking $\theta_{1,2} = \pi / 2$ in Eqs.~\eqref{eq:gam_lam_def}, \eqref{eq:pol_vec_def} and \eqref{eq:gam_lam_vs_gam_+x}. With these setup, we specifically compute the correlation functions satisfying the isosceles triangle condition ($x_1 = x_2$), or equivalently, the equal-time condition ($z_1 = z_2$), in which the angles $\phi_{1,2}$ and the separation $x_{12}$ are solely specified by the opening angle $\Theta$: $\phi_1 = (\pi - \Theta) / 2$, $\phi_2 = (\pi + \Theta) / 2$ and $x_{12} = x_1 \sqrt{2(1 - \cos\Theta)}$. Thus, the correlation function at a given redshift is expressed as a single-variate function of $\Theta$ or $x_{12}$.

Our numerical results of the correlation functions under the specific setup are shown in the top panel of Fig.~\ref{fig:xi_xiPP_per_x12}, where the function $\xi_{0 \pm 2}^{u\gamma}$ is multiplied by the factor of $50$ for ease of comparison. Overall, behaviors of the correlation function exhibit the trend similar to the TripoSH coefficients. As expected from Fig.~\ref{fig:xi_TripoSHcoeff_per_x12},  the correlation function $\xi_{0 \pm 2}^{\delta \gamma}$ has the largest signal in redshift space, though it is slightly reduced in real space. By contrast, the function $\xi_{0 \pm 2}^{u\gamma}$ is found to be rather suppressed compared to the other correlations. Mathematically, this is because both of the directional cosines $\hat{x}_{12} \cdot \hat{x}_1$ and $\hat{x}_{12} \cdot \hat{x}_2$ remain small for a relevant range of $\Theta$, and this results in a suppressed amplitude in the basis functions, ${}_{0 \pm 2}{\cal X}_{112}$ and ${}_{0 \pm 2}{\cal X}_{312}$. In fact, in the PP limit (i.e., $\hat{x}_{12} \cdot \hat{x}_1 = \hat{x}_{12} \cdot \hat{x}_2 = 0$), the correlation function $\xi_{0 \pm 2}^{u\gamma}$ becomes exactly zero \citep{Okumura:2019ned}. Regardless of the suppressed amplitude, we have the non-zero signal for the $u\gamma$ correlation. This solely comes from the wide-angle effects. The difference in magnitude between $\xi_{\pm 2 \pm 2}^{\gamma\gamma}$ and $\xi_{\pm 2 \mp 2}^{\gamma\gamma}$ arises from the inequality between ${}_{\pm 2 \pm 2}{\cal X}_{\ell \ell_1\ell_2}$ and ${}_{\pm 2 \mp 2}{\cal X}_{\ell \ell_1\ell_2}$.

The bottom panel of Fig.~\ref{fig:xi_xiPP_per_x12} plots the ratios of our wide-angle correlations to those computed in the PP limit (dubbed as $\xi_{\lambda_1 \lambda_2 \rm PP}^{X_1 X_2}$) as in \citet{Okumura:2019ned}. We find that the PP approximation works very well at small $\Theta$, while the deviation exceeds $\sim 10\%$ and becomes significant for $\Theta \gtrsim 30^\circ - 50^\circ$, as we have similarly seen in the $\delta \delta$ case.

\section{Conclusions}\label{sec:con}

We have explored the wide-angle effects on the galaxy IA statistics in the three-dimensional configuration space for the first time. The computation of the wide-angle correlation function including the spin-2 ellipticity field has become available via the complete decomposition theorem by a newly introduced spin-weighted TripoSH basis. It is because the form of the correlation function can be transformed into rather simplified expressions involving only the one-dimensional integral, as given at Eqs.~\eqref{eq:TripoSH_Xi_lam_def}, \eqref{eq:hankel_delta_u_lam} and \eqref{eq:TripoSH_Pi_delta_u_lam_iso}. Through the spin-weighted TripoSH decomposition, uninvestigated and important features in wide-angle correlations have been revealed.

Comparison of our correlation functions with those previously computed in the PP limit, given as a function of the opening angle $\Theta$, reveals that the quantitative impact of the wide-angle effects can become large by more than $10\%$ in the density-ellipticity and ellipticity-ellipticity correlations when the opening angle reaches at $\Theta \gtrsim 30^\circ - 50^\circ$. We have also found non-zero velocity-ellipticity correlation although it exactly vanishes in the PP limit. These should be taken care of in precision measurements of the IA statistics, particularly for the purpose to probe new physics with a large-angular correlation via the decadal or more futuristic surveys.

The spin-weighted TripoSH decomposition should be applied or extended to various directions. Regarding the correlation functions of the spin-0 fields such as the density and velocity fields, general relativistic effects and the effects of cosmic isotropy violation were successfully investigated via the standard TripoSH decomposition \citep[e.g.,][]{Bertacca:2012tp,Yoo:2013zga,Shiraishi:2016wec,Shiraishi:2020pea}. There are also studies on the application to the covariance matrix computation \citep{Shiraishi:2020nnw,Shiraishi:2020pea}. The same level of high versatility is naturally expected also in the spin-weighted TripoSH decomposition, and the similar issues could be resolved. Since there is no applicable limit on the value of spin in our formalism, it is even possible to extend the analysis to higher-spin IA statistics, recently proposed in \citet{Kogai:2020vzz}, and, more broadly, two-point correlations between any spinning fields.

In this Letter, all the predictions of the IA statistics are based on the linear theory, and hence the applicability of our results to small scales is limited. The validity of our predictions has to be clarified with N-body simulations and the higher-order perturbation theory.



\section*{Acknowledgements}

MS is supported by JSPS KAKENHI Grant Nos.~JP19K14718 and JP20H05859. AT acknowledges the support from MEXT/JSPS KAKENHI Grant Nos.~JP16H03977, JP17H06359, and JP20H05861. AT was also supported by JST AIP Acceleration Research Grant No.~JP20317829, Japan. TO acknowledges support from the Ministry of Science and Technology of Taiwan under Grant No.~MOST 109-2112-M-001-027- and the Career Development Award, Academia Sinica (AS-CDA-108-M02) for the period of 2019 to 2023. KA is supported by JSPS KAKENHI Grant Nos.~JP19J12254 and JP19H00677. MS and KA also acknowledge the Center for Computational Astrophysics, National Astronomical Observatory of Japan, for providing the computing resources of Cray XC50.



\section*{Data Availability Statements}

The data underlying this article are available in the article.



\bibliographystyle{mnras}
\bibliography{paper} 

\begin{thebibliography}{}
\makeatletter
\relax
\def\mn@urlcharsother{\let\do\@makeother \do\$\do\&\do\#\do\^\do\_\do\%\do\~}
\def\mn@doi{\begingroup\mn@urlcharsother \@ifnextchar [ {\mn@doi@}
  {\mn@doi@[]}}
\def\mn@doi@[#1]#2{\def\@tempa{#1}\ifx\@tempa\@empty \href
  {http://dx.doi.org/#2} {doi:#2}\else \href {http://dx.doi.org/#2} {#1}\fi
  \endgroup}
\def\mn@eprint#1#2{\mn@eprint@#1:#2::\@nil}
\def\mn@eprint@arXiv#1{\href {http://arxiv.org/abs/#1} {{\tt arXiv:#1}}}
\def\mn@eprint@dblp#1{\href {http://dblp.uni-trier.de/rec/bibtex/#1.xml}
  {dblp:#1}}
\def\mn@eprint@#1:#2:#3:#4\@nil{\def\@tempa {#1}\def\@tempb {#2}\def\@tempc
  {#3}\ifx \@tempc \@empty \let \@tempc \@tempb \let \@tempb \@tempa \fi \ifx
  \@tempb \@empty \def\@tempb {arXiv}\fi \@ifundefined
  {mn@eprint@\@tempb}{\@tempb:\@tempc}{\expandafter \expandafter \csname
  mn@eprint@\@tempb\endcsname \expandafter{\@tempc}}}

\bibitem[\protect\citeauthoryear{{Akitsu}, {Kurita}, {Nishimichi}, {Takada}  \&
  {Tanaka}}{{Akitsu} et~al.}{2020a}]{Akitsu:2020jvx}
{Akitsu} K.,  {Kurita} T.,  {Nishimichi} T.,  {Takada} M.,   {Tanaka} S.,
  2020a, arXiv e-prints, \href
  {https://ui.adsabs.harvard.edu/abs/2020arXiv200703670A} {arXiv:2007.03670}

\bibitem[\protect\citeauthoryear{{Akitsu}, {Li}  \& {Okumura}}{{Akitsu}
  et~al.}{2020b}]{Akitsu:2020fpg}
{Akitsu} K.,  {Li} Y.,   {Okumura} T.,  2020b, arXiv e-prints, \href
  {https://ui.adsabs.harvard.edu/abs/2020arXiv201106584A} {arXiv:2011.06584}

\bibitem[\protect\citeauthoryear{Bertacca, Maartens, Raccanelli  \&
  Clarkson}{Bertacca et~al.}{2012}]{Bertacca:2012tp}
Bertacca D.,  Maartens R.,  Raccanelli A.,   Clarkson C.,  2012, \mn@doi [JCAP]
  {10.1088/1475-7516/2012/10/025}, 10, 025

\bibitem[\protect\citeauthoryear{Biagetti \& Orlando}{Biagetti \&
  Orlando}{2020}]{Biagetti:2020lpx}
Biagetti M.,  Orlando G.,  2020, \mn@doi [JCAP]
  {10.1088/1475-7516/2020/07/005}, 07, 005

\bibitem[\protect\citeauthoryear{Castorina \& White}{Castorina \&
  White}{2018}]{Castorina:2017inr}
Castorina E.,  White M.,  2018, \mn@doi [Mon. Not. Roy. Astron. Soc.]
  {10.1093/mnras/sty410}, 476, 4403

\bibitem[\protect\citeauthoryear{Castorina \& White}{Castorina \&
  White}{2020}]{Castorina:2019hyr}
Castorina E.,  White M.,  2020, \mn@doi [Mon. Not. Roy. Astron. Soc.]
  {10.1093/mnras/staa2129}, 499, 893

\bibitem[\protect\citeauthoryear{Catelan, Kamionkowski  \& Blandford}{Catelan
  et~al.}{2001}]{Catelan:2000vm}
Catelan P.,  Kamionkowski M.,   Blandford R.~D.,  2001, \mn@doi [Mon. Not. Roy.
  Astron. Soc.] {10.1046/j.1365-8711.2001.04105.x}, 320, L7

\bibitem[\protect\citeauthoryear{Chisari \& Dvorkin}{Chisari \&
  Dvorkin}{2013}]{Chisari:2013dda}
Chisari N.~E.,  Dvorkin C.,  2013, \mn@doi [JCAP]
  {10.1088/1475-7516/2013/12/029}, 12, 029

\bibitem[\protect\citeauthoryear{Chisari, Dvorkin, Schmidt  \& Spergel}{Chisari
  et~al.}{2016}]{Chisari:2016xki}
Chisari N.~E.,  Dvorkin C.,  Schmidt F.,   Spergel D.,  2016, \mn@doi [Phys.
  Rev. D] {10.1103/PhysRevD.94.123507}, 94, 123507

\bibitem[\protect\citeauthoryear{Crittenden, Natarajan, Pen  \&
  Theuns}{Crittenden et~al.}{2002}]{Crittenden:2000au}
Crittenden R.~G.,  Natarajan P.,  Pen U.-L.,   Theuns T.,  2002, \mn@doi
  [Astrophys. J.] {10.1086/338838}, 568, 20

\bibitem[\protect\citeauthoryear{Croft \& Metzler}{Croft \&
  Metzler}{2000}]{Croft:2000gz}
Croft R.~A.,  Metzler C.~A.,  2000, \mn@doi [Astrophys. J.] {10.1086/317856},
  545, 561

\bibitem[\protect\citeauthoryear{Hamilton}{Hamilton}{1997}]{Hamilton:1997zq}
Hamilton A.,  1997, in {Ringberg Workshop on Large Scale Structure}.
  (\mn@eprint {arXiv} {astro-ph/9708102}),
  \mn@doi{10.1007/978-94-011-4960-0\_17}

\bibitem[\protect\citeauthoryear{Heavens, Refregier  \& Heymans}{Heavens
  et~al.}{2000}]{Heavens:2000ad}
Heavens A.,  Refregier A.,   Heymans C.,  2000, \mn@doi [Mon. Not. Roy. Astron.
  Soc.] {10.1046/j.1365-8711.2000.03907.x}, 319, 649

\bibitem[\protect\citeauthoryear{Hirata \& Seljak}{Hirata \&
  Seljak}{2004}]{Hirata:2004gc}
Hirata C.~M.,  Seljak U.,  2004, \mn@doi [Phys. Rev. D]
  {10.1103/PhysRevD.82.049901}, 70, 063526

\bibitem[\protect\citeauthoryear{{Hirata}, {Mandelbaum}, {Ishak}, {Seljak},
  {Nichol}, {Pimbblet}, {Ross}  \& {Wake}}{{Hirata} et~al.}{2007}]{Hirata:2007}
{Hirata} C.~M.,  {Mandelbaum} R.,  {Ishak} M.,  {Seljak} U.,  {Nichol} R.,
  {Pimbblet} K.~A.,  {Ross} N.~P.,   {Wake} D.,  2007, \mn@doi [\mnras]
  {10.1111/j.1365-2966.2007.12312.x}, \href
  {http://adsabs.harvard.edu/abs/2007MNRAS.381.1197H} {381, 1197}

\bibitem[\protect\citeauthoryear{Kogai, Matsubara, Nishizawa  \& Urakawa}{Kogai
  et~al.}{2018}]{Kogai:2018nse}
Kogai K.,  Matsubara T.,  Nishizawa A.~J.,   Urakawa Y.,  2018, \mn@doi [JCAP]
  {10.1088/1475-7516/2018/08/014}, 08, 014

\bibitem[\protect\citeauthoryear{{Kogai}, {Akitsu}, {Schmidt}  \&
  {Urakawa}}{{Kogai} et~al.}{2020}]{Kogai:2020vzz}
{Kogai} K.,  {Akitsu} K.,  {Schmidt} F.,   {Urakawa} Y.,  2020, arXiv e-prints,
  \href {https://ui.adsabs.harvard.edu/abs/2020arXiv200905517K} {arXiv:2009.05517}

\bibitem[\protect\citeauthoryear{{Mandelbaum}, {Hirata}, {Ishak}, {Seljak}  \&
  {Brinkmann}}{{Mandelbaum} et~al.}{2006}]{Mandelbaum:2006}
{Mandelbaum} R.,  {Hirata} C.~M.,  {Ishak} M.,  {Seljak} U.,   {Brinkmann} J.,
  2006, \mn@doi [\mnras] {10.1111/j.1365-2966.2005.09946.x}, \href
  {http://adsabs.harvard.edu/abs/2006MNRAS.367..611M} {367, 611}

\bibitem[\protect\citeauthoryear{{Okumura} \& {Jing}}{{Okumura} \&
  {Jing}}{2009}]{Okumura:2009a}
{Okumura} T.,  {Jing} Y.~P.,  2009, \mn@doi [\apjl]
  {10.1088/0004-637X/694/1/L83}, \href
  {http://adsabs.harvard.edu/abs/2009ApJ...694L..83O} {694, L83}

\bibitem[\protect\citeauthoryear{Okumura \& Taruya}{Okumura \&
  Taruya}{2020}]{Okumura:2019ned}
Okumura T.,  Taruya A.,  2020, \mn@doi [Mon. Not. Roy. Astron. Soc.]
  {10.1093/mnrasl/slaa024}, 493, L124

\bibitem[\protect\citeauthoryear{{Okumura}, {Jing}  \& {Li}}{{Okumura}
  et~al.}{2009}]{Okumura:2009}
{Okumura} T.,  {Jing} Y.~P.,   {Li} C.,  2009, \mn@doi [\apj]
  {10.1088/0004-637X/694/1/214}, \href
  {http://adsabs.harvard.edu/abs/2009ApJ...694..214O} {694, 214}

\bibitem[\protect\citeauthoryear{Okumura, Taruya  \& Nishimichi}{Okumura
  et~al.}{2019}]{Okumura:2019ozd}
Okumura T.,  Taruya A.,   Nishimichi T.,  2019, \mn@doi [Phys. Rev. D]
  {10.1103/PhysRevD.100.103507}, 100, 103507

\bibitem[\protect\citeauthoryear{{Okumura}, {Taruya}  \&
  {Nishimichi}}{{Okumura} et~al.}{2020}]{Okumura:2020a}
{Okumura} T.,  {Taruya} A.,   {Nishimichi} T.,  2020, \mn@doi [\mnras]
  {10.1093/mnras/staa718}, \href
  {https://ui.adsabs.harvard.edu/abs/2020MNRAS.494..694O} {494, 694}

\bibitem[\protect\citeauthoryear{Papai \& Szapudi}{Papai \&
  Szapudi}{2008}]{Papai:2008bd}
Papai P.,  Szapudi I.,  2008, \mn@doi [Mon. Not. Roy. Astron. Soc.]
  {10.1111/j.1365-2966.2008.13572.x}, 389, 292

\bibitem[\protect\citeauthoryear{Schmidt \& Jeong}{Schmidt \&
  Jeong}{2012}]{Schmidt:2012ne}
Schmidt F.,  Jeong D.,  2012, \mn@doi [Phys. Rev. D]
  {10.1103/PhysRevD.86.083527}, 86, 083527

\bibitem[\protect\citeauthoryear{Schmidt, Chisari  \& Dvorkin}{Schmidt
  et~al.}{2015}]{Schmidt:2015xka}
Schmidt F.,  Chisari N.~E.,   Dvorkin C.,  2015, \mn@doi [JCAP]
  {10.1088/1475-7516/2015/10/032}, 10, 032

\bibitem[\protect\citeauthoryear{Shiraishi, Nitta, Yokoyama, Ichiki  \&
  Takahashi}{Shiraishi et~al.}{2011}]{Shiraishi:2010kd}
Shiraishi M.,  Nitta D.,  Yokoyama S.,  Ichiki K.,   Takahashi K.,  2011,
  \mn@doi [Prog. Theor. Phys.] {10.1143/PTP.125.795}, 125, 795

\bibitem[\protect\citeauthoryear{Shiraishi, Sugiyama  \& Okumura}{Shiraishi
  et~al.}{2017}]{Shiraishi:2016wec}
Shiraishi M.,  Sugiyama N.~S.,   Okumura T.,  2017, \mn@doi [Phys. Rev. D]
  {10.1103/PhysRevD.95.063508}, 95, 063508

\bibitem[\protect\citeauthoryear{{Shiraishi}, {Okumura}  \&
  {Akitsu}}{{Shiraishi} et~al.}{2020a}]{Shiraishi:2020pea}
{Shiraishi} M.,  {Okumura} T.,   {Akitsu} K.,  2020a, arXiv e-prints, \href
  {https://ui.adsabs.harvard.edu/abs/2020arXiv200904355S} {arXiv:2009.04355}

\bibitem[\protect\citeauthoryear{Shiraishi, Okumura, Sugiyama  \&
  Akitsu}{Shiraishi et~al.}{2020b}]{Shiraishi:2020nnw}
Shiraishi M.,  Okumura T.,  Sugiyama N.~S.,   Akitsu K.,  2020b, \mn@doi [Mon.
  Not. Roy. Astron. Soc.] {10.1093/mnrasl/slaa132}, 498, L77

\bibitem[\protect\citeauthoryear{Szalay, Matsubara  \& Landy}{Szalay
  et~al.}{1998}]{Szalay:1997cc}
Szalay A.~S.,  Matsubara T.,   Landy S.~D.,  1998, \mn@doi [Astrophys. J.
  Lett.] {10.1086/311293}, 498, L1

\bibitem[\protect\citeauthoryear{Szapudi}{Szapudi}{2004}]{Szapudi:2004gh}
Szapudi I.,  2004, \mn@doi [Astrophys. J.] {10.1086/423168}, 614, 51

\bibitem[\protect\citeauthoryear{Taruya \& Okumura}{Taruya \&
  Okumura}{2020}]{Taruya:2020tdi}
Taruya A.,  Okumura T.,  2020, \mn@doi [Astrophys. J. Lett.]
  {10.3847/2041-8213/ab7934}, 891, L42

\bibitem[\protect\citeauthoryear{Taruya, Saga, Breton, Rasera  \&
  Fujita}{Taruya et~al.}{2020}]{Taruya:2019xsf}
Taruya A.,  Saga S.,  Breton M.-A.,  Rasera Y.,   Fujita T.,  2020, \mn@doi
  [Mon. Not. Roy. Astron. Soc.] {10.1093/mnras/stz3272}, 491, 4162

\bibitem[\protect\citeauthoryear{Varshalovich, Moskalev  \&
  Khersonsky}{Varshalovich et~al.}{1988}]{Varshalovich:1988ye}
Varshalovich D.~A.,  Moskalev A.~N.,   Khersonsky V.~K.,  1988, {Quantum Theory
  of Angular Momentum: Irreducible Tensors, Spherical Harmonics, Vector
  Coupling Coefficients, 3nj Symbols}.
World Scientific, Singapore

\bibitem[\protect\citeauthoryear{{Vlah}, {Chisari}  \& {Schmidt}}{{Vlah}
  et~al.}{2020a}]{Vlah:2020ovg}
{Vlah} Z.,  {Chisari} N.~E.,   {Schmidt} F.,  2020a, arXiv e-prints, \href
  {https://ui.adsabs.harvard.edu/abs/2020arXiv201204114V} {arXiv:2012.04114}

\bibitem[\protect\citeauthoryear{Vlah, Chisari  \& Schmidt}{Vlah
  et~al.}{2020b}]{Vlah:2019byq}
Vlah Z.,  Chisari N.~E.,   Schmidt F.,  2020b, \mn@doi [JCAP]
  {10.1088/1475-7516/2020/01/025}, 01, 025

\bibitem[\protect\citeauthoryear{Yoo \& Seljak}{Yoo \&
  Seljak}{2015}]{Yoo:2013zga}
Yoo J.,  Seljak U.,  2015, \mn@doi [Mon. Not. Roy. Astron. Soc.]
  {10.1093/mnras/stu2491}, 447, 1789

\makeatother
\end{thebibliography}


\appendix
\section{Angular correlation function} \label{appen:Cl}

When the field ${}_{\lambda} X$ is dealt with on the celestial sphere via the spherical harmonic expansion:
\begin{align}
  {}_{\lambda} X({\bf x}) = \sum_{\ell m} {}_{\lambda} a_{\ell m}^{X} \, {}_{\lambda} Y_{\ell m} (\hat{x}) ,
\end{align}
the correlation function of ${}_{\lambda} a_{\ell m}^{X}$ obeys
\begin{align}
  \Braket{{}_{\lambda_1} a_{\ell_1 m_1}^{X_1} {}_{\lambda_2} a_{\ell_2 m_2}^{X_2}} = {}_{\lambda_1 \lambda_2} C_{\ell_1}^{X_1 X_2} (-1)^{m_1} \delta_{\ell_1, \ell_2} \delta_{m_1, -m_2} ,
\end{align}
and is therefore related to $\xi_{\lambda_1 \lambda_2}^{X_1 X_2}$ as
\begin{align} 
  \xi_{\lambda_1 \lambda_2}^{X_1 X_2}({\bf x}_{12}, \hat{x}_1, \hat{x}_2)
  = \sum_{\ell m}
  {}_{\lambda_1 \lambda_2} C_{\ell}^{X_1 X_2} (-1)^{\lambda_2}
  {}_{\lambda_1} Y_{\ell m} (\hat{x}_1) {}_{-\lambda_2} Y_{\ell m}^* (\hat{x}_2) .
\end{align}
In the analysis of the IA statistics, the E/B-mode decomposition,
\begin{align}
  {}_{0} a_{\ell m}^E &\equiv - \frac{1}{2} \left({}_{+2} a_{\ell m}^\gamma + {}_{-2} a_{\ell m}^\gamma \right) , \\
  {}_{0} a_{\ell m}^B &\equiv - \frac{1}{2i} \left({}_{+2} a_{\ell m}^\gamma - {}_{-2} a_{\ell m}^\gamma \right),
\end{align}
has also been conventionally utilized \citep{Crittenden:2000au}. Since ${}_{+2} a_{\ell m}^\gamma = {}_{-2} a_{\ell m}^\gamma$ in our case, the following relations hold:  
\begin{align}
  {}_{00}C_\ell^{\delta E} &= - \, {}_{0 \pm 2} C_{\ell}^{\delta \gamma} , \\
  {}_{00}C_\ell^{u E} &= - \, {}_{0 \pm 2} C_{\ell}^{u \gamma} , \\
  {}_{00}C_\ell^{EE} &= {}_{\pm 2 \pm 2} C_{\ell}^{\gamma \gamma} = {}_{\pm 2 \mp 2} C_{\ell}^{\gamma \gamma} , \\
  {}_{00}C_\ell^{\delta B} &= {}_{00}C_\ell^{u B} = {}_{00}C_\ell^{EB} = {}_{00}C_\ell^{BB} = 0 .
\end{align}


\bsp	
\label{lastpage}
\end{document}